# On the Physical Origin of the Anomalous Magnetic Moment of Electron

N. B. MANDACHE

National Institute for Laser, Plasma and Radiation Physics, Lab.22, P.O. Box. MG-36, 077125 Magurele -Bucharest, Romania, (electronic address: mandache@infim.ro)

**Abstract:** A simple physical insight into the origin of the magnetic moment anomaly of electron is presented. This approach is based on the assumption that the electromagnetic mass of the electron due to the electric field generated by electron charge in the exterior of the sphere of radius $\lambda_e/2$ ( where $\lambda_e$ is the electron Compton wavelength) does not contribute to the magnetic moment of the electron. This explanation is compatible with the well-known quantum electrodynamics approach. A formula is derived, which is similar to that obtained by quantum electrodynamics calculus of one loop contribution to anomalous part of the magnetic moment.

The anomalous magnetic moment of the electron is calculated with very high accuracy in quantum electrodynamics (QED) and agrees with the experimentally measured value to more than 10 significant figures [1]. On the other hand there is no simple physical insight into the origin of the moment anomaly [2, 3].

It is suggested [2] that a way to qualitatively understand this effect may be found by remembering that the magnetic moment of the Dirac electron $e\hbar/2m$ is due to circular currents of the radius $\lambda_e = h/mc$. The zero-point oscillations of the electromagnetic field and the current fluctuations induced in the "vacuum" influence these currents to a certain extent and thus cause the slight change of the magnetic moment [2]. The first calculus of the anomalous magnetic moment of the electron was made by Schwinger [4] who obtained a new formula for the magnetic moment in very good agreement with the experiment: $\frac{e\hbar}{2m}(1+\alpha/2\pi)$, where $\alpha$ is the fine structure constant.

The physical interpretation of the magnetic moment $e\hbar/2m$ of the Dirac electron, originating in a circular motion of radius equal to the reduced Compton wavelength $\left(\lambdabar_e = \hbar/mc\right)$ is largely treated in literature [5- 8]. At the non-relativistic limit the electron does not behave like a point charge but like a distribution of charge and current, extended over a region of radius ~ $\hbar/mc$.

On the other hand, the mass of the electron has a spatial distribution extended towards infinite. Indeed there is definite experimental evidence that some of the mass of a charged particle is electromagnetic in origin [9]. It is meaningless to separate a charge, in particular that of electron, from its Coulomb field. Neither can be observed without the other; a charge at rest is always surrounded by a coulomb field and conversely every Coulomb field has a source [10]. Coulomb field means electrostatic energy and consequently electrostatic mass for electron. For instance the electrostatic energy (mass) of an electron in the region outside of a sphere of radius r (where $r > \lambdabar_e$) is equal to $e^2/2r$ [9].

Based on the above discussion we propose the following mechanism to explain the anomaly in the magnetic moment: only the mass of the electron which is inside the sphere of radius ~ $\lambda_e$ participates at the "dynamics" (circular currents) that generates the magnetic moment. The electron mass outside the sphere of radius ~ $\lambda_e$ does not contribute to the magnetic moment of the electron. This mass, $\Delta m \cdot c^2$ is equal to the electrostatic energy in the electric field generated by the electron charge in the exterior of the sphere of radius ~ $\lambda_e$. Instead of $e\hbar/2m$, the new expression of the magnetic moment is:

$$\mu = \frac{e\hbar}{2(m - \Delta m)} = \frac{e\hbar}{2m\left(1 - \frac{\Delta m}{m}\right)} = \frac{e\hbar}{2m\left(1 - \frac{e^2}{2k\lambda_e mc^2}\right)} \qquad (1)$$

where $k\lambda_e$ is the radius of separation between the two regions of electron mass distribution, k is a numerical coefficient and $\Delta m \cdot c^2 = e^2/2k\lambda_e$. From the above considerations it results that the radius $k\lambda_e$ has a value between $\lambdabar_e$ and $\lambda_e$. For k=1/2 one obtains:

$$\frac{\dfrac{e^2}{2k\lambda_e}}{mc^2} = \frac{\dfrac{e^2}{\lambda_e}}{mc^2} = \frac{\dfrac{e^2}{\lambdabar_e}}{2\pi \cdot mc^2} = \frac{r_e}{2\pi \cdot \lambdabar_e} = \frac{\alpha}{2\pi} \tag{2}$$

where $r_e$ is the classical electron radius ($r_e = e^2/mc^2$) and $\lambdabar_e = r_e \alpha^{-1}$.

Replacing (2) in relation (1) it results:

$$\mu = \frac{e\hbar}{2m\left(1 - \dfrac{\alpha}{2\pi}\right)} \cong \frac{e\hbar}{2m}\left(1 + \frac{\alpha}{2\pi} + ...\right) \tag{3}$$

The relation (3) is similar to that derived by Schwinger.

The attempts to evaluate radiative corrections to electron phenomena have encountered difficulties due to divergences attributable to self-energy and vacuum polarization effects [4]. To avoid these difficulties, at moderate energies, the Hamiltonian of current hole theory electrodynamics was transformed to exhibit explicitly the logarithmically divergent self-energy of a free electron. The electromagnetic self-energy of a free electron can be ascribed to an electromagnetic mass, which must be added to the mechanical mass of the electron. The only meaningful statements of the theory involve the sum of these two masses, which is the experimental mass of a free electron. Further in [4] it is underlined that however the transformation of the Hamiltonian is based on the hypothesis of a weak interaction between matter and radiation, which means that the electromagnetic mass must be a small correction $\sim \alpha \cdot m_0$ to the "mechanical" mass $m_0$.

The above description of the electron mass structure (distribution) given in [4], in particular the weak interaction between the small electromagnetic mass of the electron and the mechanical mass of the electron, supports our assumption that the electromagnetic mass outside the sphere of radius $\lambda_e/2$ does not contribute to the magnetic moment of the electron. From relation (2) it results that the electromagnetic mass outside the sphere of radius $\lambda_e/2$ is equal to $\dfrac{\alpha}{2\pi} \cdot m$.